\documentclass[sigconf]{acmart}

\usepackage{balance}
\usepackage{algorithmic}
\usepackage{subfigure}
\usepackage{bm}
\usepackage{float}
\usepackage[ruled,linesnumbered]{algorithm2e}
\usepackage{graphicx}  

\AtBeginDocument{%
  }

\setcopyright{acmcopyright}
\copyrightyear{2022}
\acmYear{2022}

 \acmConference[UbiComp/ISWC '22 Adjunct]{the 2022 ACM International Joint Conference on Pervasive and Ubiquitous Computing and Proceedings of the 2022 ACM International Symposium on Wearable Computers (UbiComp/ISWC '22 Adjunct), September 15, 2022, Virtual Event, Global.}{September 15, 2022}{Virtual Event, Global}
 \acmPrice{15.00}
 \acmISBN{978-1-4503-XXXX-X/18/06}

\begin{document}

\title{Power and Interference Control for VLC-Based UDN: A Reinforcement Learning Approach}


\author{Xiao Tang}
\affiliation{%
  \institution{tangx@stu.xmu.edu.cn}
  \institution{Xiamen University}
  \city{Xiamen}
  \state{Fujian}
  \country{China}}

  \author{Sicong Liu}
  \authornote{Corresponding author.}
  \affiliation{%
    \institution{liusc@xmu.edu.cn}
    \institution{Xiamen University}
    \city{Xiamen}
    \state{Fujian}
    \country{China}}


\begin{abstract}
  Visible light communication (VLC) has been widely applied as a promising solution for modern short range communication.
    When it comes to the deployment of LED arrays in VLC networks, the emerging ultra-dense network (UDN) technology can be adopted to expand the VLC network's capacity. However, the problem of inter-cell interference (ICI) mitigation and efficient power control in the VLC-based UDN is still a critical challenge.
    To this end, a reinforcement learning (RL) based VLC UDN architecture is devised in this paper. The deployment of the cells is optimized via spatial reuse to mitigate ICI. An RL-based algorithm is proposed to dynamically optimize the policy of power and interference control, maximizing the system utility in the complicated and dynamic environment.
    Simulation results demonstrate the superiority of the proposed scheme, it increase the system utility and achievable data rate while reducing the energy consumption and ICI, which outperforms the benchmark scheme.
\end{abstract}





\maketitle

\section{Introduction}
{\color{black}Visible light communication (VLC) has drawn great attention due to its potential in wideband unregulated spectrum and low implementation cost \cite{t1,txuma}. Meanwhile, the electromagnetic-interfer-ence-free characteristic of VLC provides an alternative of wireless coverage in some RF-constrained scenarios.

Ultra-dense network (UDN) is an up-and-coming scheme in next-generation mobile communications \cite{txiao}. Higher system data rates can be achieved by dense spatial reuse of wireless spectrum using UDN.
Thus, it is very promising to apply the UDN technology in VLC networks to improve its performance. However, the system performance might be impacted seriously due to the inter-cell interference (ICI) of dense deployment light emitting diode (LED) access points (APs) in the VLC-based UDN \cite{t2}.
}

Numerous schemes have been reported in literature and patents to mitigate ICI for VLC networks. A spatial multiplexing method is proposed, all cells share a broadcast channel $f_{0}$, and the sub-carrier set is spatially reused to enable communication with less interference \cite{t3}. Grouping based scheme is often adopted to against ICI, the cells with the largest mutual interference are dynamically selected as a group to cooperatively transmit information \cite{t4,t5}.
A heuristic algorithm of power redistribution between the interference-constrained sub-carriers is proposed to solve the problem of ICI \cite{t6}. In state-of-the-art research, some receiver-side processing schemes, such as a differential optical detection scheme \cite{t7} and an optimized angle diversity receiver \cite{t8}, these scheme address the challenge of ICI in VLC systems while increasing the complexity of hardware implementation.
With the introduction of UDN in VLC systems, the ICI will get even worse and dominant due to the dense deployment of VLC APs. ICI severely impacts the system performance, bringing about a great challenge remains to be properly addressed.

Recently, the application of reinforcement learning (RL) has been a common tool for high-dimensional optimization problems, which provides a possible solution for optimal decision making in complicated and dynamic systems \cite{txiao1}.
For instance, a RL-based VLC beamforming control scheme is proposed to realize the optimal beamforming strategy against eavesdroppers \cite{t10}. An online dual-time scale power distribution algorithm is proposed, using multi-agent Q-learning \cite{t11}. In addition, UAVs can use RL-based anti-jamming transmission schemes to adaptively counter external interference attacks \cite{t12}.
The environment of VLC-based UDN is composed of APs, user equipment (UE), and time-varying channel states and ICI. We desperately need a way to manage the highly complicated and dynamic system.

To improve the performance of the VLC-based UDN by mitigating the ICI, a Q-learning-based scheme is proposed in the RL framework to determine the optimal power control strategy and to maximize the system utility, the proposed scheme improves the energy efficiency by minimizing the transmitting power, while the ICI is restricted and the acceptable achievable data rate is guaranteed.
Specifically, the channel state information (CSI), the signal-to-interference-and-noise power ratio (SINR), the density of UEs constitute the state space for the RL-based process while the transmitting power of the APs is regarded as the action space. Actions are determined based on the proposed RL-based algorithm, it is capable of converging to the optimal action that achieves utility maximization based on the state information \cite{t13}.

The rest of this paper is organized as follows: {\color{black} the model of the VLC-based UDN is introduced in Section II. The proposed RL-based power and interference control scheme is presented in Section III. Section IV reports the simulation results and discussions, and conclusions of this work are drawn in Section V.}

\section{System Model and Cell Deployment of the VLC-Based UDN}
A UDN system with densely deployed VLC-based APs, i.e., LED arrays, is considered to provide broadband transmission access for multiple VLC-based UEs. In general, the VLC transmission channel is composed of line-of-sight (LoS) and non-line-of-sight (NLoS) paths. It has been revealed that the strength of the LoS path in the VLC channel is dominant compared with the NLoS paths \cite{t131}. Thus, for simplicity of presentation, only the LoS path is considered in terms of calculating the SINR and the interference in the VLC transmission.
In fact, our generic algorithm proposed in this paper can also be extended to different VLC channel models, including those considering NLoS paths. According to the widely adopted Lambertian VLC channel model \cite{t132}, the channel gain between the AP and the $n$-th UE is given as
\begin{equation}\label{sec4:h}\small
    {h_n} = \frac{{(m + 1){A_{\rm{R}}}}}{{2\pi d_n^2}}{\cos ^m}(\phi )\cos (\theta ){\rm{rect}}(\theta ),
\end{equation}
where $A_{\mathrm{R}}$ and $d_{n}$ are the effective detection area of the photodiode (PD) and the distance between VLC AP and the $n$-th active user, respectively. Angle parameters of $\phi $, $\theta $ and ${\phi _{1/2}}$ denote the irradiance angle, the incidence angle, and the half-intensity radiation angle, respectively. The parameter $m = -1/\log _{2}\left(\cos \phi_{1 / 2}\right)$  denotes the Lambertian order. The rectangular function ${\rm{rect}}(\theta )$ is given by
\begin{equation}\label{sec3:g}\small
    {\rm{rect}}(\theta ) = \left\{ {\begin{array}{*{20}{c}}
        {1,{\rm{|}}\theta {\rm{|}} \le {\theta _{{\rm{FOV}}},}}\\
        {0,{\rm{|}}\theta {\rm{|}} > {\theta _{{\rm{FOV}}},}}
        \end{array}} \right.
\end{equation}
where ${\theta _{{\rm{FOV}}}}$ is the field-of-view (FOV) angle of the PD.
\begin{figure}[htbp]
    \centering
    \subfigure[Two-spectrum-block Mode]{
    \label{fig:q1}
    \includegraphics[width=3.09 in]{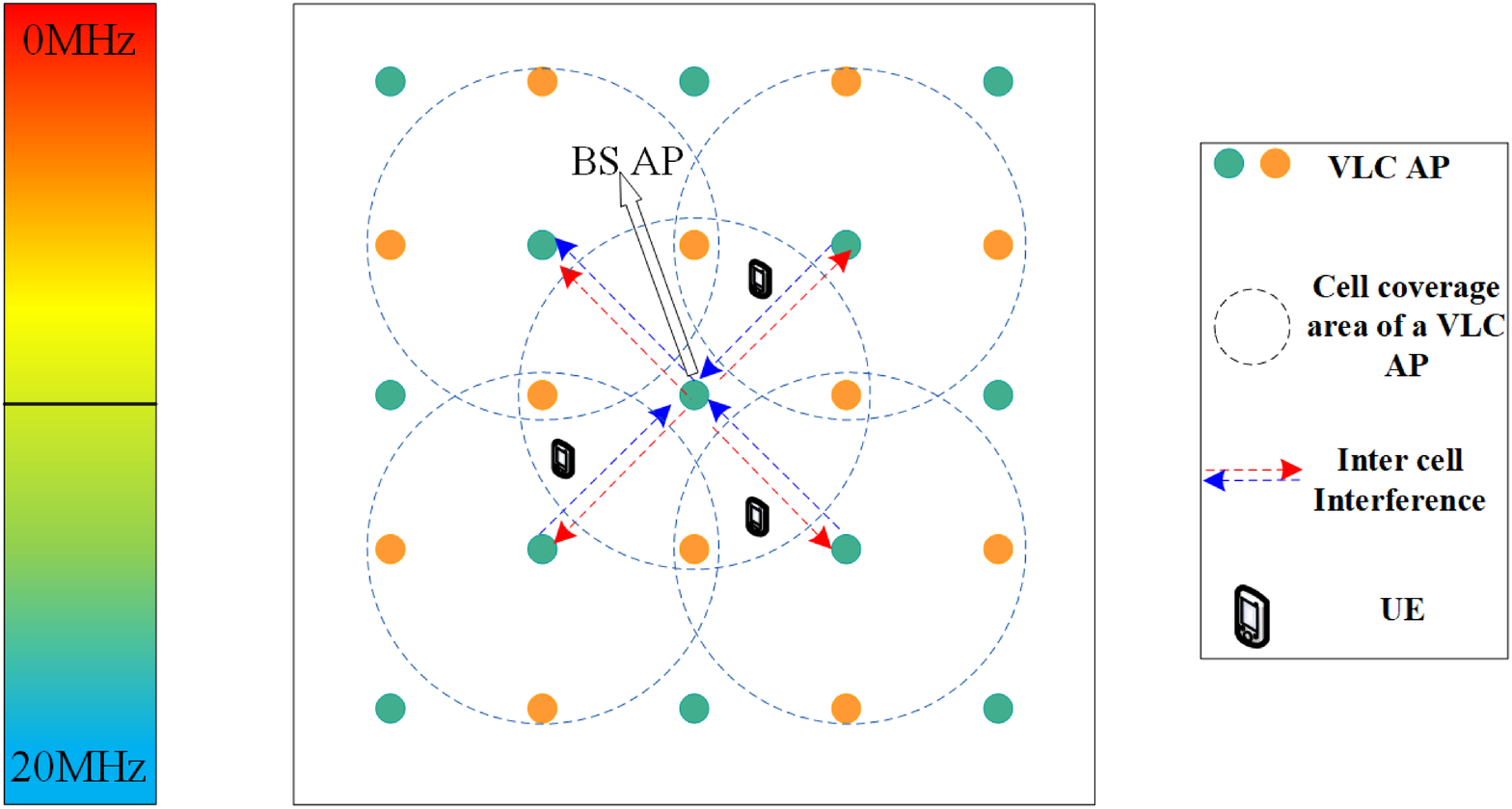}}
    \subfigure[Four-spectrum-block Mode]{
    \label{fig:q2}
    \includegraphics[width=3.09 in]{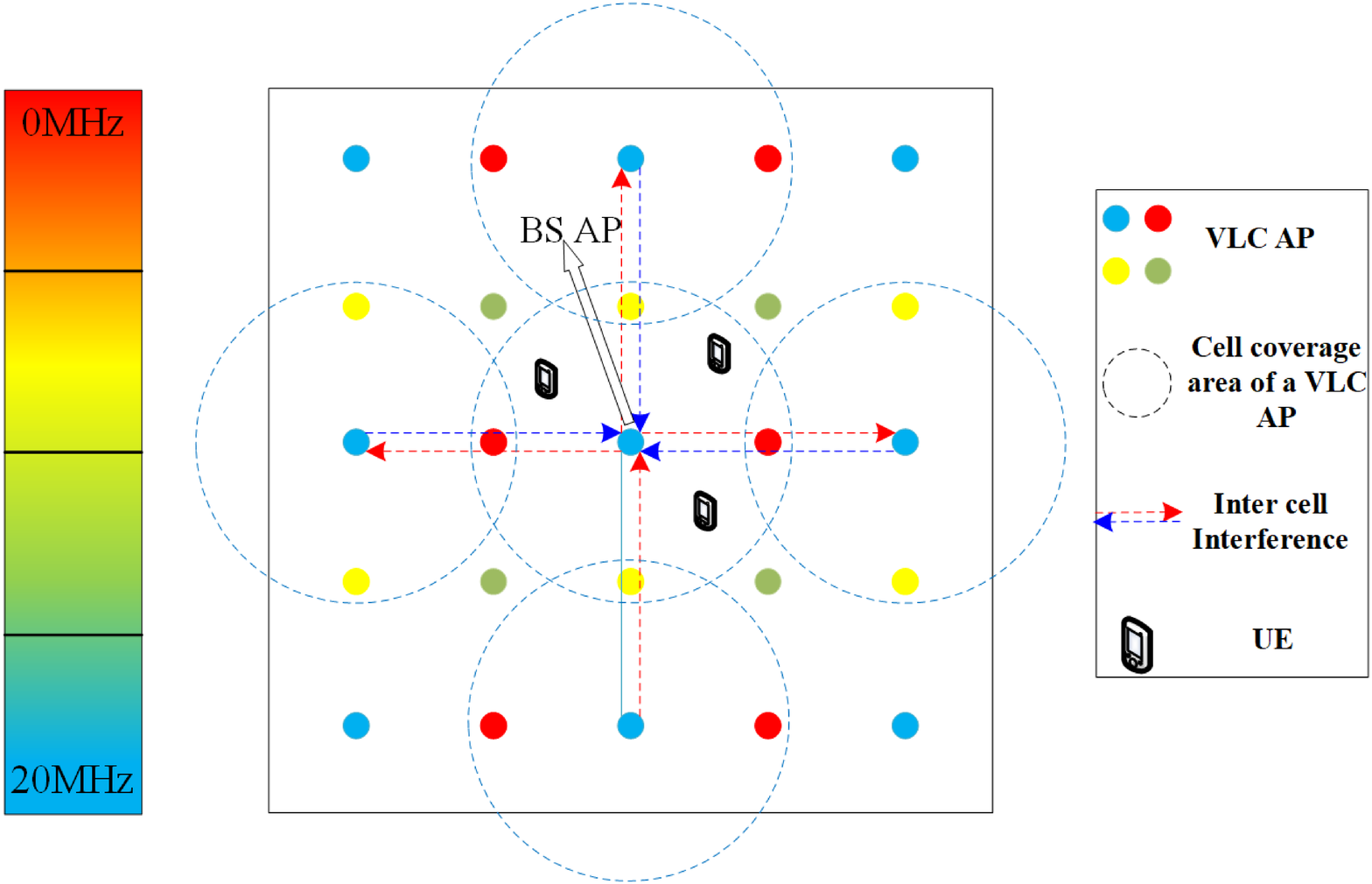}}
    \caption{Cell deployment layouts in the considered VLC-based UDN: (a) Two-spectrum-block mode; (b) Four-spectrum-block mode.}
    \label{figq}
\end{figure}

In the VLC-based UDN, adjacent cells are assigned to different frequency bands to implement spatial reuse of the spectrum. The available bandwidth of the VLC transmission employing intensity modulation and direct detection (IM/DD) is divided into a certain number of orthogonal spectrum blocks, and the number of the spectrum blocks is a crucial question to be considered in the system.
For one thing, if the UDN is deployed with only a few blocks, the distance between the cells using the same spectrum blocks would be smaller, which leads to severe ICI. On the other hand, the capacity of each AP would be limited due to the dense partitioning of the total spectrum. In order to pursue a greater throughput of the UDN, a certain level of ICI can be tolerated, and measures can be taken to eliminate the ICI.

According to most of the existing literature, the shape of the VLC cells is a square rather than a cellular shape adopted in radio frequency cellular networks \cite{t14,t15,t16,t17}. The VLC service is only an extra functionality of LED arrays, layout of them suitable for lighting is considered with priority. We assume that the VLC model operates under a power line communication (PLC) system as in \cite{tplc}, and the PLC can act as a link between the indoor LEDs and the outdoor network.
As shown in Fig. \ref{figq}, considering an indoor VLC scenario in which the overall VLC bandwidth is divided into several spectrum blocks, the topology of the cell is in the same way as the cellular network. The total bandwidth is assumed to be 20 MHz. The dots with different colors represent the APs occupying different spectrum blocks. Usually, the mobile users prefer to connect to the nearest AP since it is most likely to provide the best quality-of-service.
In Fig. \ref{figq}, the layout of a VLC-based UDN with 25 APs is illustrated, and in fact there can be more APs in practice. The distance between APs and the FOV angle of the LED arrays in the UDN can be optimized according to the specific details of the communication scenarios.

\begin{figure}[h]
  \centering
  \includegraphics[width=3.3 in]{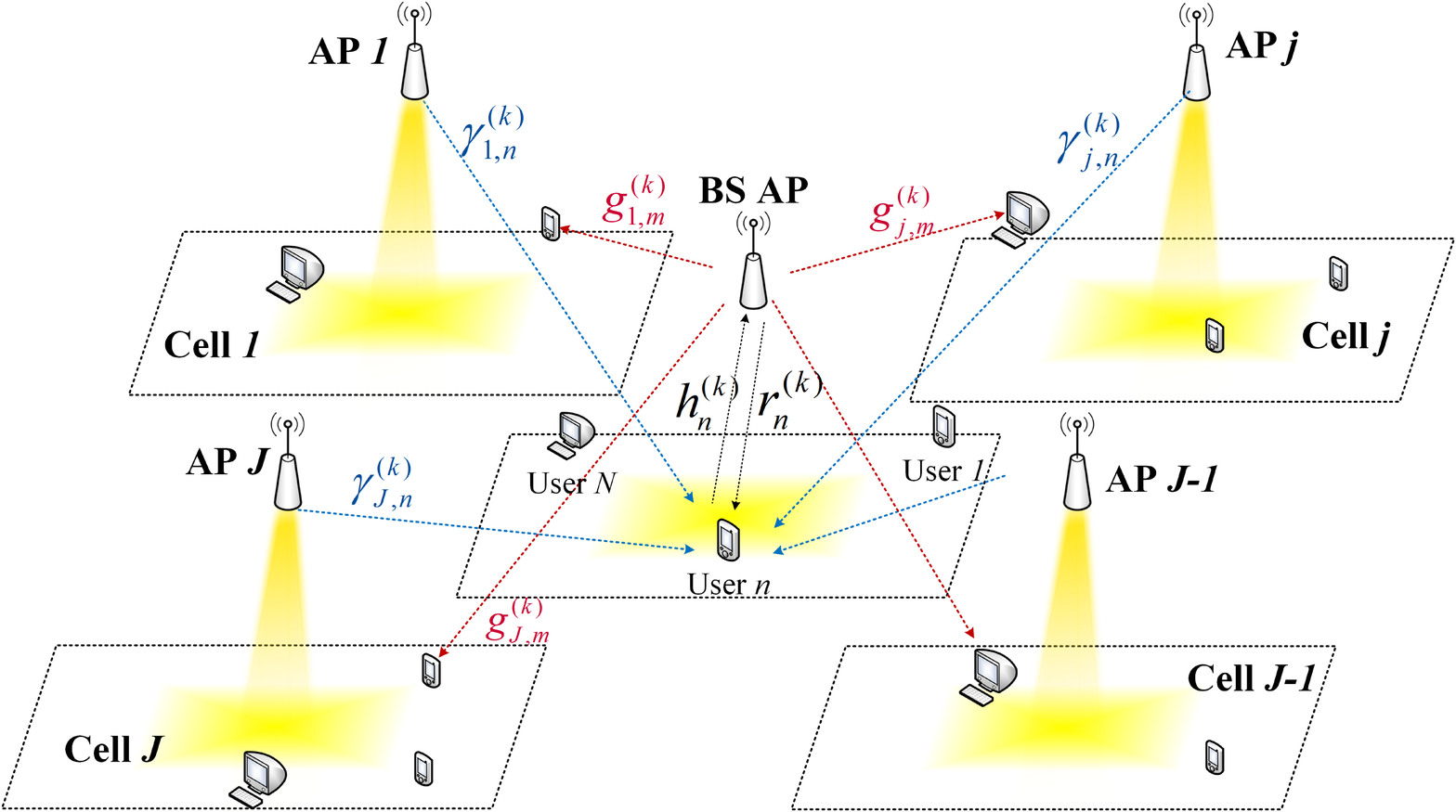} 
  \caption{The central BS AP serves $N$ mobile UEs in the central cell, and determines its actions of transmit power $x_{n}^{(k)}$, maximizing the cell throughput and meanwhile to save the energy consumption and to mitigate the ICI to the neighboring $J$ cells at time slot $k$.
  The estimated SINR $r_{n}^{(k)}$ of the $n^{th}$ UE is sent to the central BS AP as feedback information.}
  \label{system1}
\end{figure}

As described above, the dense deployment of cells in the VLC-based UDN leads to serious ICI. In the two-spectrum-block mode as shown in Fig. \ref{fig:q1}, once a user is close to the edge of the cell, the SINR decays rapidly due to severe ICI from more and closer cells. In contrast, the four-spectrum-block mode in Fig. \ref{fig:q2} has solved this problem by increasing the distance between the interfering APs occupying the same spectrum blocks. It constrains the performance differences between central users and peripheral users within the same cell.

Considering the performance of the UEs in the central cell served by the main base station (BS) AP, as illustrated in Fig. \ref{system1}, interactions of cells are depicted and the four-spectrum-block mode is adopted in our subsequent simulations. The index of time slots is denoted by $k$. $h_{n}^{(k)}$ denotes the downlink channel gain of $n$-th UE. The interference channel gain from the $j$-th neighboring AP to the $n$-th UE in the BS AP is denoted by $\gamma_{j, n}^{(k)}$, and $g_{j,m}^{(k)}$ is the channel gain of the interference from the BS AP to the $m$-th UE in the $j$-th neighboring AP.
Thus, the SINR of the $n$-th UE in the BS AP at time slot $k$ is calculated as
\begin{equation}\label{sec3:sinr}\small
    \zeta_{n}^{(k)}=\frac{\left(\eta x_{n}^{(k)} h_{n}^{(k)}\right)}{W_{n} N_{0}+\sum_{j=1}^{J}\left(\eta x_{j,n}^{(k)} \gamma_{j, n}^{(k)}\right)},
\end{equation}
where $x_{n}^{(k)}$ is the transmitted optical power from the BS AP to the $n$-th UE served by the BS, which is proportional to the amplitude of the electrical information signal, i.e. the current driving the LED, in commonly applied IM/DD based VLC transmission. $\eta $ is the opto-electro-converting efficiency of the PD. $x_{j,n}^{(k)}$ is the power of the interference from the $j$-th neighboring AP to the $n$-th UE in the BS AP. $N_{0}$ is the power spectral density (PSD) of the background noise.
$J$ represents the number of the neighboring APs around the BS AP. With $N$ being the number of UEs therein, the bandwidth is uniformly and orthogonally allocated to the UEs in the BS AP, i.e. ${W_n} = W/N$, where $W$ and $W_n$ denote the total available bandwidth of the BS AP and the bandwidth allocated to the $n$-th UE, respectively. Therefore, the interference $\chi^{(k)}$ to neighboring APs and the achievable data rate $r_n^{(k)}$ of $n$-th UE can be measured as
\begin{equation}\label{sec4:interf}\small
    \chi^{(k)}  = \sum\limits_{n = 1}^N {\sum\limits_{j = 1}^J {\sum\limits_{m = 1}^{M_j^{(k)}} {{{(\eta x_n^{(k)}{g_{j,m}^{(k)}})}}} } },
\end{equation}
\begin{equation}\label{sec4:datarate}\small
    r_n^{(k)} = {W_n}{\log _2}(1 + \zeta _n^{(k)}),
\end{equation}
{\color{black}where $M_j^{(k)}$ in (4) denotes the number of active UEs at time slot $k$ in the $j$-th neighboring AP.}

\section{Reinforcement-Learning-Based Power and Interference Control Scheme in VLC-Based UDN}

In this section, the RL-based power and interference control (RPIC) in VLC-based UDN system is proposed. For the sake of better adaption to the nonstationary VLC channel due to UE movement, each AP determines the transmit power by adopting the RPIC algorithm in every time slot $k$.

As shown in Fig. \ref{system1}, the square in the center represents the coverage of the central cell served by the BS AP. The neighboring cells around the central cell will produce ICI on the UEs served by the BS AP in the central cell. The ICIs between the BS AP and the neighboring APs are represented by dashed lines with the corresponding channel gains annotated in Fig. \ref{system1}.
It is necessary for the proposed power and interference control scheme to mitigate the ICI from the BS AP to the neighboring cells with best efforts, while ensuring the communication quality-of-service for the UEs it served.

The actions, states, and utility of the proposed RPIC algorithm is described as follows. Subject to the dimming and lighting requirements of the LED arrays, the optical transmit power of the $n$-th UE denoted by $x_n^{(k)}$, has a maximum value of $\mathcal{X}_{\max }$ \cite{t18}. In order to formulate the action set to be selected in the RL-based algorithm, the feasible set of the power value is quantized into $L$ levels expressed as $\Omega  = {\{ i{{\mathcal{X} }_{\max }}/L\} _{0 \le i \le L}}$.

In time slot $k$, the state of the central cell in the VLC-based UDN is given by $\mathbf{s}^{(k)} = (\mathbf{r}^{(k-1)}, {\rho ^{(k)}, \mathbf{h}^{(k)}}) \in \textbf{S}$, includeing the achievable data rate of the UEs, i.e. $\mathbf{r}^{(k-1)} = \left( {r_1^{(k-1)},r_2^{(k-1)}...,r_N^{(k-1)}} \right)$, the CSI vector, i.e. $\mathbf{h}^{(k)} = \left( {h_1^{(k)},h_2^{(k)}...,h_N^{(k)}} \right)$, and UE density of the cell, i.e. ${\rho ^{(k)}}$. The three components in the state can be approximately measured or estimated by the UEs in the cell, and sent to the BS AP as feedback information.

The utility function $u ^{(k)}$ for the BS AP in the central cell is used to update the Q-function of the RPIC algorithm at the $k$-th time slot. $u ^{(k)}$ is composed of three factors, a proper trade-off can be achieved among these factors by adding some weights to reflect the influence of each factor on the utility. According to equation (3), (4) and (5), the utility at time slot $k$ can be denoted as
\begin{equation}\label{sec4:utility}\small
    \begin{split}
        u^{(k)}  =  &\frac{1}{N}\sum\limits_{n = 1}^N {W_n}{\log _2}(1 + \zeta _n^{(k)}) -\\
        &{C_\textup{E}}\sum\limits_{n = 1}^N {x_n^{(k)}}  -{C_\textup{I}}\sum\limits_{n = 1}^N {\sum\limits_{j = 1}^J {\sum\limits_{m = 1}^{M_{j}^{(k)}} {{{(\eta {x_n^{(k)}}{g_{j,m}^{(k)}})},}} } }\\
    \end{split}
\end{equation}
\begin{algorithm}[h]
  \caption{RL-based Power and Interference Control (RPIC).}
  $\mathbf{Initialize}$ $N$, $\alpha $, $\beta $, and $u^{(0)} = 0$ \\
  $Q\left( {{\bf{s}},{\bf{x}}} \right) \leftarrow 0,\forall {\bf{s}} \in S,{\bf{x}} \in {\Omega ^N}$ \\
  Randomly choose an action $\mathbf{x}^{(0)} \in {\Omega ^N}$ \\
  \For{$k = 1:{K_{\max }}$}{
       Obtain the throughput vector feedback $\mathbf{r}^{(k-1)}$ \\
       Estimate the UE density at current slot $\rho ^{(k)}$ \\
       Estimate the CSI vector $\mathbf{h}^{(k)}$  \\
       Formulate the current state $\mathbf{s}^{(k)} = (\mathbf{r}^{(k-1)}, {\rho ^{(k)}, \mathbf{h}^{(k)}})$ \\
       Choose the action vector $\mathbf{x}^{(k)}$ via (7) \\
       The BS AP conducts the selected action $\mathbf{x}^{(k)}$ \\
       Measure the feedback throughput vector $\mathbf{r}^{(k)}$ \\
       Deriving the system utility $u^{(k)}$ using (6) \\
       Update the Q-function by (8)  \\}
\end{algorithm}

where $C_\textup{E}$ and $C_\textup{I}$ are the weights for the energy consumption and the ICI, respectively. Among the factors to the right of equation (6), the first term is the average achievable data rate of the UEs in the central cell, which is intended to be increased for a larger VLC network throughput. The second term reflects the impact of the energy consumption on the utility due to the transmit power of the BS AP serving the UEs in this cell, aims to drive the algorithm to learn an energy-efficient strategy. The last term reflects the negative impact of the ICI on the system utility. Appropriate weights of $C_\textup{E}$ and $C_\textup{I}$ can improve the system performance in a proper tradeoff that balances the throughput, energy consumption, and ICI.


\begin{figure*}[htbp]
  \centering
  \subfigure[Utility of the VLC-based UDN]{
  \label{fig:f1}
  \includegraphics[width=2.6 in]{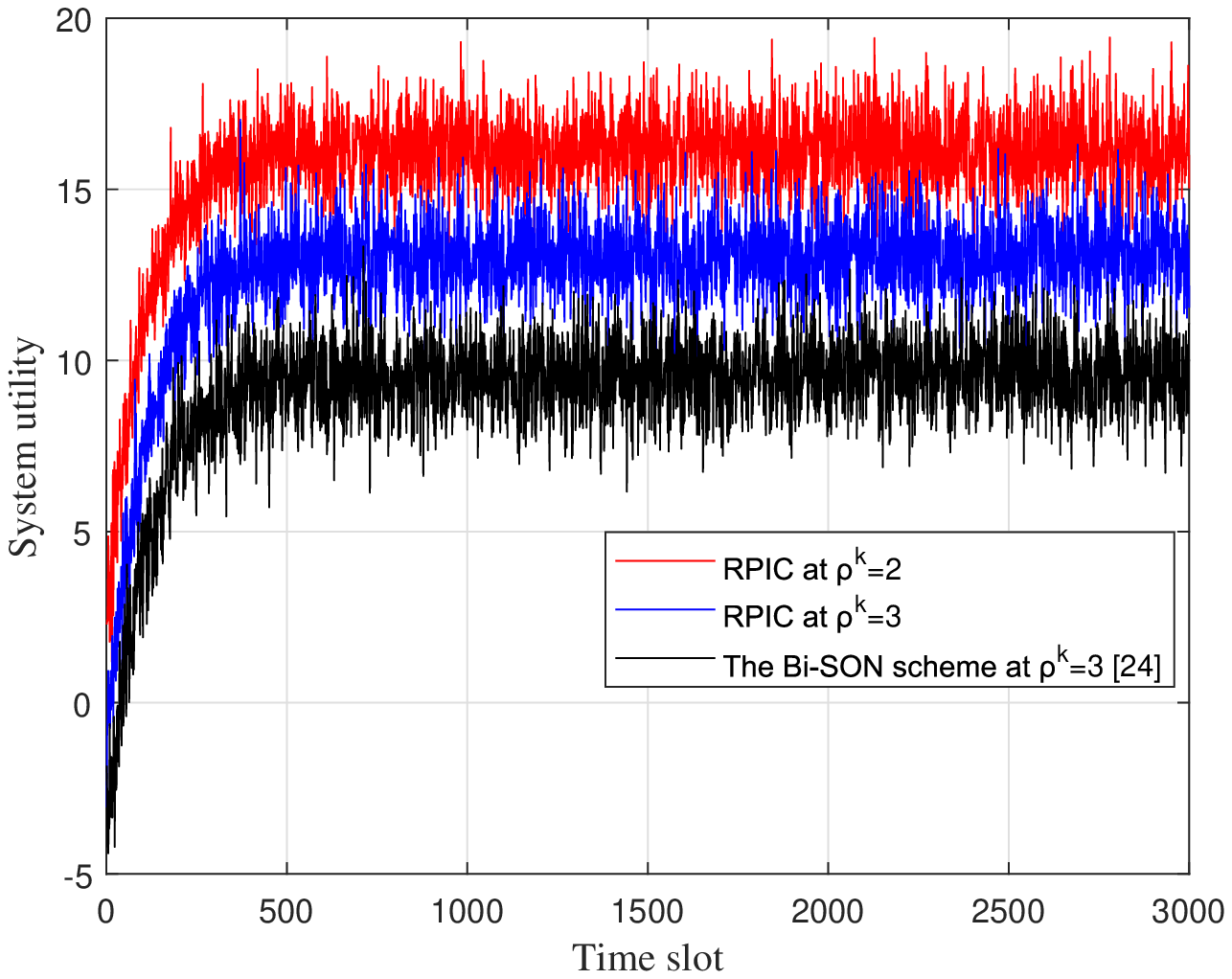}}
  \subfigure[Average throughput of the UEs in the UDN]{
  \label{fig:f2}
  \includegraphics[width=2.6 in]{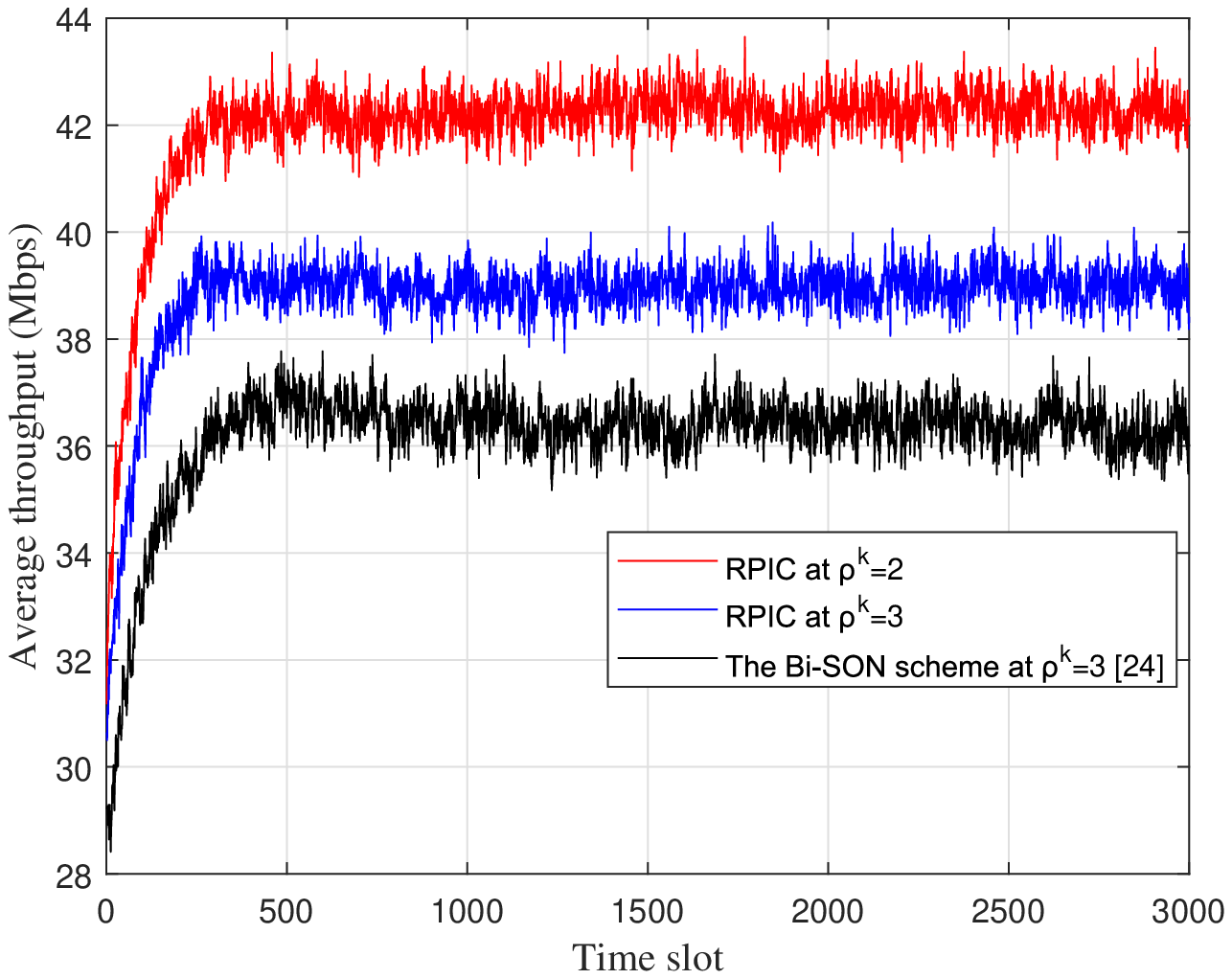}}
  \subfigure[Energy consumption of the BS AP]{
  \label{fig:f3}
  \includegraphics[width=2.6 in]{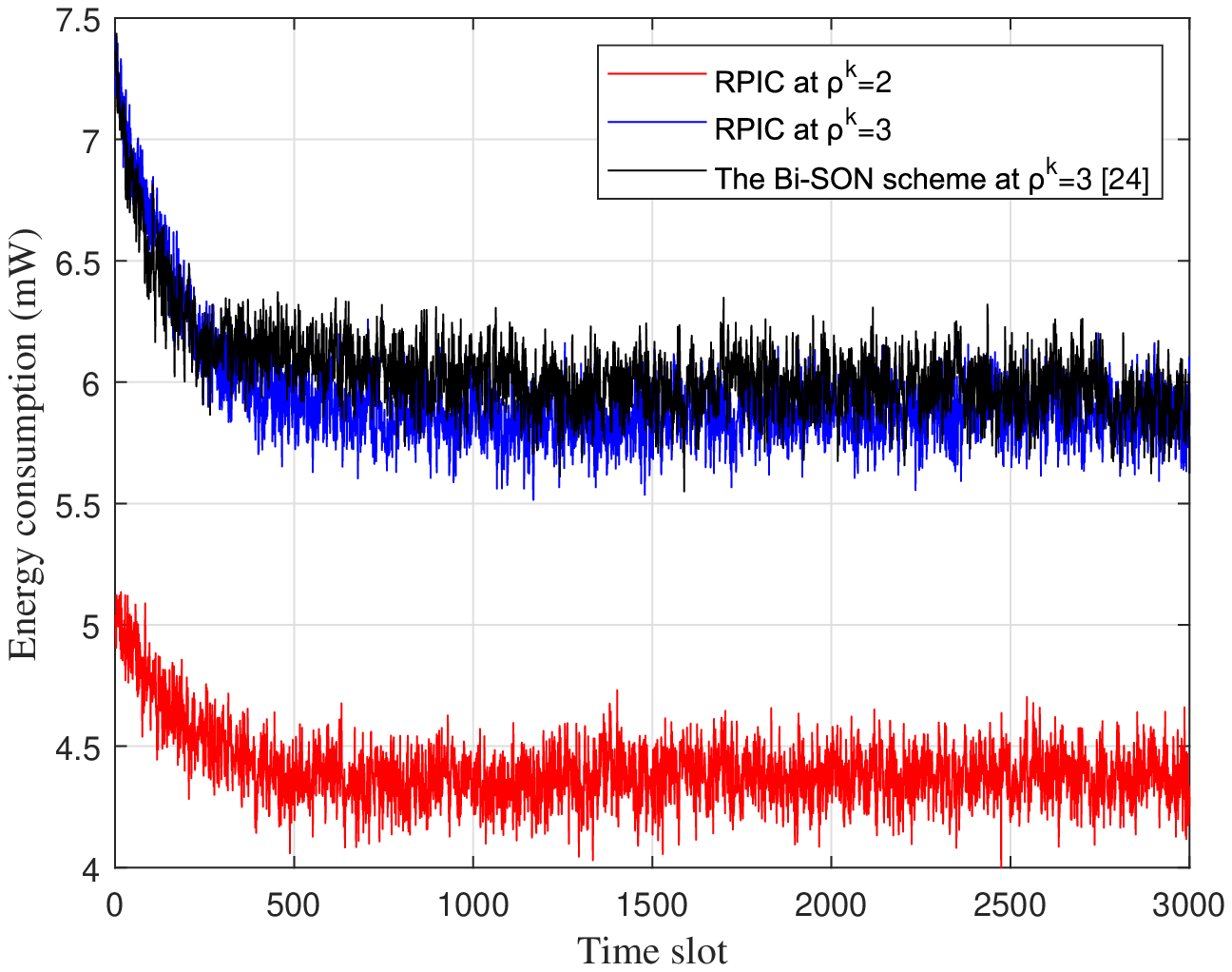}}
  \subfigure[ICI to neighboring cells]{
  \label{fig:f4}
  \includegraphics[width=2.6 in]{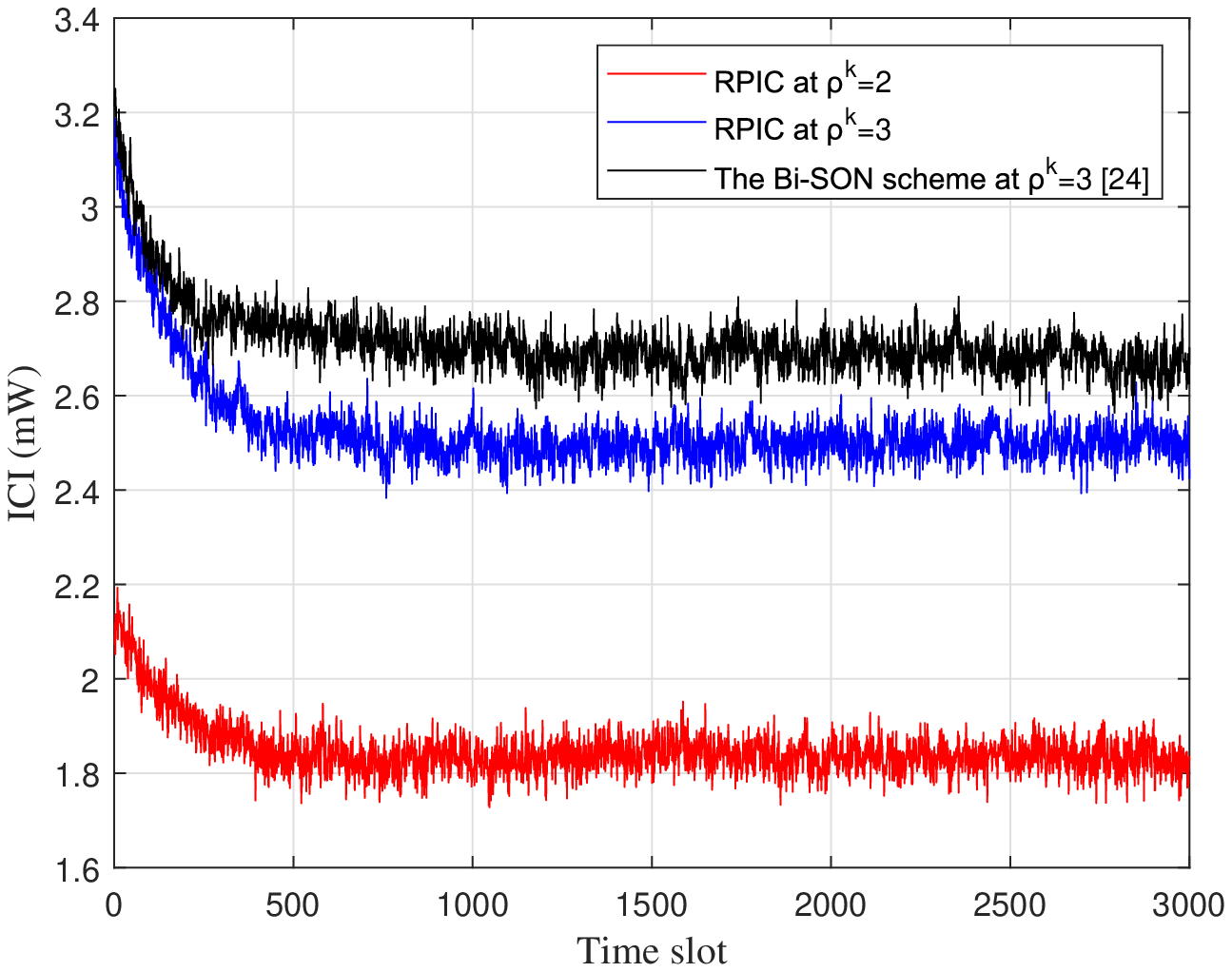}}
  \caption{Performance of the proposed RPIC scheme and the benchmark Bi-SON scheme with different UE density: (a) VLC-based UDN utility, (b) Average throughput of UEs, (c) Energy consumption, and (d) ICI.}
  \label{figq1}
\end{figure*}

The details of the RPIC algorithm are shown in \textbf{Algorithm 1}. The Q-function $Q\left(\mathbf{s}^{(k)},\mathbf{x}^{(k)}\right)$ is a policy function for BS AP to determine the transmit power to users, where $\mathbf{s}^{(k)} \in \textbf{S}$ represents the current state, and ${{\bf{x}}^{(k)}} = {\left[ {x_n^{(k)}} \right]_{1 \le n \le N}},{\rm{ }}x_n^{(k)} \in {\Omega}$ is the action vector representing the transmit powers allocated to the UEs.
The value of the Q-function, i.e. $Q\left(\mathbf{s}^{(k)},\mathbf{x}^{(k)}\right)$, indicates the reward obtained by choosing the action of $\mathbf{x}^{(k)}$ in a the current state of $\mathbf{s}^{(k)}$. A higher value of Q-function indicates the feasibility to choose the action $\mathbf{x}^{(k)}$ in the current state $\mathbf{s}^{(k)}$. The Bellman iteration equation updates the value of Q-function in every slot according to \textbf{Algorithm 1}.

The action $\mathbf{x}^{(k)}$ selected for time slot $k$ is determined by the $\epsilon$-greedy criterion as given by:
\begin{align}\small
    \Pr \left({\mathbf{x}^{(k)}} = {\hat{\mathbf{x}}}\right)=
    \begin {cases}
    1-\epsilon, & {\hat{\mathbf{x}}}= \arg\max\limits_{\mathbf{x}} Q\left(\mathbf{s}^{(k)},{\mathbf{x}}\right)\\
    \frac{\epsilon}{|\Omega ^N|-1},&\textrm{o.w}.
    \end{cases} \label {greedy}
\end{align}
where $\hat{\mathbf{x}}$ is the action vector that can maximize the Q-value at the current state $\mathbf{s}^{(k)}$, and $\mathbf{x}=\left\{\left[x_{n}\right]_{1 \leq n \leq N} \mid x_{n} \in \Omega\right\}$ is any action vector including $N$ actions selected from the action set.
The "state-action" pair will be evaluated by (6), and the selected pair's value $Q\left(\mathbf{s}^{(k)},\mathbf{x}^{(k)}\right)$ in the Q-function will be updated by the iterative Bellman equation below:
\begin{equation}\label{sec4:utility}\small
  \begin{split}
    Q(\mathbf{s}^{(k)},\mathbf{x}^{(k)})
    &\leftarrow(1-\alpha )Q(\mathbf{s}^{(k)},\mathbf{x}^{(k)})\\
      &+\alpha (u^{(k)}+\beta \max_{\mathbf{x}} Q(\mathbf{s}^{(k+1)},\mathbf{x})).\\
  \end{split}
\end{equation}

\section{SIMULATION RESULTS and Discussions}

In this section, simulations are conducted to evaluate the performance of the proposed RPIC scheme for the VLC-based UDN system, comparing with the popular benchmark scheme for power control in dense cellular networks, i.e., Bi-SON \cite{tb}. As shown in Fig. \ref{system1}, considering an indoor environment with the area of $10\times10\times3\  \textrm{m}^3$, where $5\times5$ VLC APs are uniformly spaced and deployed at the ceiling.
The central BS AP is impacted by the ICI from the neighboring cells around it. Due to the limited FOV angle of the LED arrays, the ICI between the BS AP and the adjacent neighboring cells are taken into consideration.
Assuming the movement routines of all UEs follow a random way point model \cite{t20}, and the destination and velocity are randomly generated. The density $\rho ^{(k)}$ is simplified to the number of users in a single cell, which is defined as:
\begin{equation}\label{sec4:utility}\small
  \rho^{(k)} = N^{(k)}.
\end{equation}
To quantitatively evaluate the optimization of the proposed RPIC scheme, we will set $\rho^{(k)}$ to keep its value constant in an episode to observe the convergence, i.e. $N^{(k)}= N$, and then use the four factors in (6) as evaluation metrics.
\begin{table}[b]
    \renewcommand{\arraystretch}{1.2}
    \caption{Geometric parameters of VLC-baesd UDN}
    \vspace{-0.25cm}  
    \setlength{\abovecaptionskip}{-0.20cm}   
    \setlength{\belowcaptionskip}{-0.1cm}   
    \centering
    \small
    \begin{tabular}{l l}
    \hline
    \qquad $\textbf{Parameters and Corresponding Value}$ \\
    \hline
    Semi-angle of half luminous intensity & $60^\circ$ \\
    Distance between APs  & $2\ \textrm{m}$  \\
    Radius of AP coverage & $2.1\ \textrm{m}$  \\
    Room height  & $3\ \textrm{m}$ \\
    FOV angle  & $70^{\circ}$ \\
    Cell size  & $2\times2\ \textrm{m}^2$ \\
    Height of UE  & $1\ \textrm{m}$ \\
    \hline
    \end{tabular}\label{table1}
\end{table}

The configuration of LED arrays are set up as follows. Each VLC AP has an available bandwidth of $W=20$ MHz, while the actual equivalent bandwidth is 10MHz due to the Hermitian symmetry before IFFT in IM/DD OFDM-based VLC transmission \cite{t19}. The optical transmit power is quantized into $L=5$ discrete levels with the maximum value of ${{\mathcal{X} }_{{\rm{max}}}}=4$ mW. The PSD of the background noise is ${N_0} = {10^{ - 21}}$ ${{\rm{A}}^2}/$Hz. The FOV angle of the PD is ${\theta _{{\rm{FOV}}}}=70^{\circ}$. The detection area AP is ${A_{\rm{R}}}=1$ ${\rm{c}}{{\rm{m}}^2}$.
The semi-angle of half luminous intensity is ${\phi _{1/2}}=60^{\circ}$, and the opto-electro-converting efficiency of the PD is $\eta =0.54$ A/W. The remaining parameters are listed in Table I.

The RPIC scheme is configured with the maximum number of iterations of ${K_{\max }}=3000$. In the early stage of the process, the smart agent, i.e. the BS AP, randomly selects a state and an action in the first 20 time slots. The current state, the action, together with the obtained utility $u$ and the next state are combined to form an experience, these experiences is saved in the experience pool to be exploited to update the Q-function. The learning rate $\alpha $ and the discount rate $\beta $ are set to 0.9 and 0.3, respectively.
The parameter of $\epsilon $ gradually decreases from 0.9 to 0.1. The simulation results reported for each time slot are calculated by averaging over 1000 runs.

The performance of the proposed RPIC scheme and the benchmark scheme Bi-SON \cite{tb} is reported in Fig. \ref{figq1}. Specifically, Fig. \ref{fig:f1} and Fig. \ref{fig:f2} present the utility and throughput of the proposed RPIC and the benchmark scheme with different UE density of $\rho ^{(k)} $. As shown by Fig. \ref{fig:f1}, with the increase of the UE density $ \rho ^{(k)} $, the overall utility of the system decreases due to more crowded users and thus more severe ICI.
Moreover, the proposed RPIC scheme outperforms the benchmark scheme of Bi-SON \cite{tb} at all time slots. The proposed RPIC scheme has a better capability to mitigate ICI and can obtain a larger throughput as shown in Fig. \ref{fig:f2} and Fig. \ref{fig:f4}.

Specifically, let us consider the case with $\rho ^{(k)} =3$. The RPIC scheme can converge to an average data rate of 42.1 Mbps after 300 frames, this value is 8.33$\%$ higher than that of the benchmark Bi-SON. Meanwhile, the energy consumption and the ICI are mitigated via RPIC scheme. Using RPIC, the energy consumption and the ICI are reduced by 13.8$\%$ and 17.3$\%$ than those at the initial state of learning as shown in Fig. \ref{fig:f3} and Fig. \ref{fig:f4}.
Moreover, the ICI of RPIC is 14.9$\%$ smaller than the value of Bi-SON with the same energy consumption. The proposed RPIC scheme's superior performance has verified that the RL-based mechanism can quickly adapt the policy to non-stationary VLC channels to achieve a better system utility and a good tradeoff between system throughput, energy efficiency, and interference immunity.

\section{Conclusion}

{\color{black}In this paper, a novel ultra-dense network architecture with efficient spatial spectrum reuse for the indoor VLC transmission system is presented, which significantly improves the system performance in terms of achievable average data rate, energy efficiency, and inter-cell interference mitigation.
We propose a reinforcement-learning-based algorithm called RPIC to adaptively and intelligently control the transmit power for the users, this algorithm aims to reduce the inter-cell interference while saving energy and guaranteeing the throughput.}
{\color{black}Simulation results have verified the superior performance of the proposed algorithm with different user density compared to the benchmark scheme, in terms of the capability to optimize for the power and interference control policy in a dynamic and non-stationary VLC transmission environment.}


\bibliographystyle{ACM-Reference-Format}
\bibliography{sample-base}

\appendix

\end{document}